\documentclass[aps, twocolumn, superscriptaddress]{revtex4}

\usepackage{amsfonts}           
\usepackage{amssymb}            
\usepackage{amsmath}            
\usepackage{comment}
\usepackage{latexsym}           
\usepackage{dcolumn}            
\usepackage[dvips]{graphicx}   
\usepackage{enumerate}
\usepackage{epstopdf}           
\usepackage{fancyhdr}           
\usepackage{hyperref}           
\usepackage{setspace}           
\usepackage{xspace}             
\usepackage{subfigure}          
\usepackage{bm}                 
\usepackage[ulem=normalem]{changes}
\newcommand{\tht}{\theta}

\newcommand{\bea}{\begin{eqnarray}}
\newcommand{\eea}{\end{eqnarray}}

\begin{document}

\title{Isostaticity at Frictional Jamming}
\date{\today}

\author{Stefanos Papanikolaou}
\affiliation{Departments of Mechanical Engineering \& Materials Science, Yale University, New Haven, Connecticut 06520}
\affiliation{Department of Physics, Yale University, New Haven, Connecticut 06520}

\author{Corey S. O'Hern}
\affiliation{Departments of Mechanical Engineering \& Materials Science, Yale University, New Haven, Connecticut 06520}
\affiliation{Department of Physics, Yale University, New Haven, Connecticut 06520}
\affiliation{Department of Applied Physics, Yale University, New Haven, Connecticut 06520}

\author{Mark D. Shattuck}
\affiliation{Benjamin Levich Institute and Physics Department, The City
College of the City University of New York, New York, New York 10031}

\begin{abstract}
Amorphous packings of frictionless, spherical particles are isostatic
at jamming onset, with the number of constraints (contacts) equal to
the number of degrees of freedom. Their structural and mechanical
properties are controlled by the interparticle contact network.  In
contrast, amorphous packings of frictional particles are typically
hyperstatic at jamming onset.  We perform extensive numerical
simulations in two dimensions of the geometrical asperity (GA) model
for static friction, to further investigate the role of isostaticity.
In the GA model, interparticle forces are obtained by summing up
purely repulsive central forces between periodically spaced circular
asperities on contacting grains. We compare the packing fraction,
contact number, mobilization distribution, and vibrational density of
states using the GA model to those generated using the Cundall-Strack
(CS) approach.  We find that static packings of frictional disks
obtained from the GA model are mechanically stable and {\it isostatic}
when we consider interactions between asperities on contacting
particles. The crossover in the structural and mechanical properties
of static packings from frictionless to frictional behavior as a
function of the static friction coefficient coincides with a change in
the type of interparticle contacts and the disappearance of a peak in
the density of vibrational modes for the GA model.  These results
emphasize that mesoscale features of the model for static friction
play an important role in determining the properties of granular
packings.
\end{abstract}

\maketitle 

Recently, intense effort has been devoted to
understanding the jamming transition of athermal frictionless spheres 
with repulsive contact
interactions~\cite{Torquato:2010ye, Liu:1998nx, OHern:2003kl,
  Van-Hecke:2010fv}.  However, physical models of granular media should
include static friction~\cite{Shafer:1996tg}. Experiments~\cite{Majmudar:2005uq, Bi:2011fk} and simulations~\cite{Song:2008kl,Silbert:2002ij,Henkes:2010qa} have
shown that amorphous frictional sphere packings can be obtained at
jamming onset over a wide contact number range $d+1 \le z \le
2d$~\cite{Edwards:1999hc, Silbert:2010mi, OHern:2003kl}, where $d$ is
the spatial dimension.  In addition, a crossover from
frictionless random close packing $\phi \simeq \phi_{\rm
  RCP}$ and $z \simeq 2d$ to frictional random loose packing
$\phi \simeq \phi_{\rm RLP}$ and $z \simeq d+1$ as the static friction
coefficient $\mu$ increases above $\mu^* \sim 0.1$ ($0.01$) in $d=2
(3)$~\cite{Silbert:2010mi}.  Moreover, a large number $N_s$ of
`sliding' contacts (with the tangential equal to the normal force times $\mu$) exists for small $\mu$, and $N_s$ decreases with
increasing $\mu$~\cite{Shundyak:2007fk,Silbert:2010mi}.  When contact-counting arguments account for
sliding contacts, frictional packings can be described as `isostatic'
with similar vibrational properties to frictionless spheres'~\cite{Henkes:2010qa}.

\begin{figure}[t!]
\centering
\includegraphics[scale=0.35]{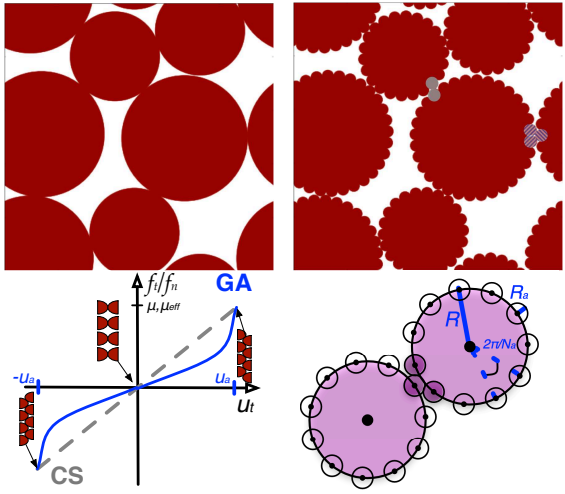}
\caption{Top: Nearly identical MS packings of $N=6$ bidisperse disks
at jamming onset from the CS (left) and GA (right) models with $\mu,
\mu_{\rm eff} \simeq 0.3$ and $\phi_J \simeq 0.78$ and $0.76$,
respectively; they possess the same $9$ interparticle contacts, and
the GA model has the isostatic number of contacting asperities
$N_c^{aa} = 3N-1=17$. (right) The central particle has five
interactions between asperities on three contacting grains.  The
solid and striped gray contacts between the central particle and its
neighbors are single and double asperity contacts,
respectively. Bottom: (left) Schematic of the ratio of tangential
and normal forces $f_t/f_n$ at constant interparticle overlap versus
the relative tangential displacement $u_t$ for the CS (dashed) and
GA (solid) models.  For CS, $f_t/f_n$ is linear with slope $k_t$,
while for GA $f_t/f_n = u_t/\sqrt{(\sigma^{aa'}_{ij}-r_{ij}^{aa'})^2
- u_t^2}$, where $r_{ij}^{aa'}$ is constant at fixed
overlap. Single (double) asperity contacts occur near $f_t/f_n = 0$
(maximal $|f_t|/f_n$).  Sliding happens when $\pm u_a = \pm \mu
f_n/k_t$ in CS, while in GA $u_a = \pm \sigma^{aa'}_{ij}/(2
\sqrt{1+1/\mu_{\rm eff}^2})$ and $f_t/f_n$ is periodic at zero
overlap. (right) Schematic of the interaction in the GA model between 
disks with radius
$R$, $N_a$ circular asperities with radius $R_a$, and angle
$2\pi/N_a$.}
\label{fig:1}
\end{figure}

In this Letter, we address several open questions: How sensitive are the structural
(dependent on particle positions) and mechanical properties
(dependent on interparticle forces) of frictional packings to the
friction model employed? What determines the static friction
coefficient $\mu^*$ that marks the crossover from frictionless to
frictional behavior for static packings? How does $D(\omega)$ for
frictional packings differ from ones of frictionless
particles with complex and anisotropic ({\it e.g.} convex and
non-convex) shapes?

Most prior studies focused on
the CS approach~\cite{Cundall:1979uq}, where static
friction is modeled by a tangential spring (with spring constant $k_t$
and restoring force $k_t u_t$, where $u_t$ is the relative tangential
displacement) when particles in contact, and the Coulomb sliding
condition holds. With the GA model we can
distinguish interparticle contacts based on which asperities
 interact and calculate $D(\omega)$ by taking
derivatives of total potential energy without making {\it ad hoc}
assumptions on sliding contacts~\cite{Henkes:2010qa}.  Prior GA models mimicking frictional interactions~\cite{torres09,
  Alonso-Marroquin:2008vn, Buchholtz:1994fk} studied
dense granular flows.

Static GA packings are mechanically stable
(MS) and {\it isostatic} when asperity interactions are considered,
independent of the effective static friction coefficient.  The
crossover as a function of the
effective friction coefficient coincides with changes in the
interaction types between asperities and the disappearance of a
strong, primarily rotational, peak in $D(\omega)$ at low
frequency. We also find that $D(\omega)$ for the GA model differs from
analogous studies for the CS case~\cite{Henkes:2010qa}.
 
\begin{figure}[t!]
\centering
\includegraphics[scale=0.44]{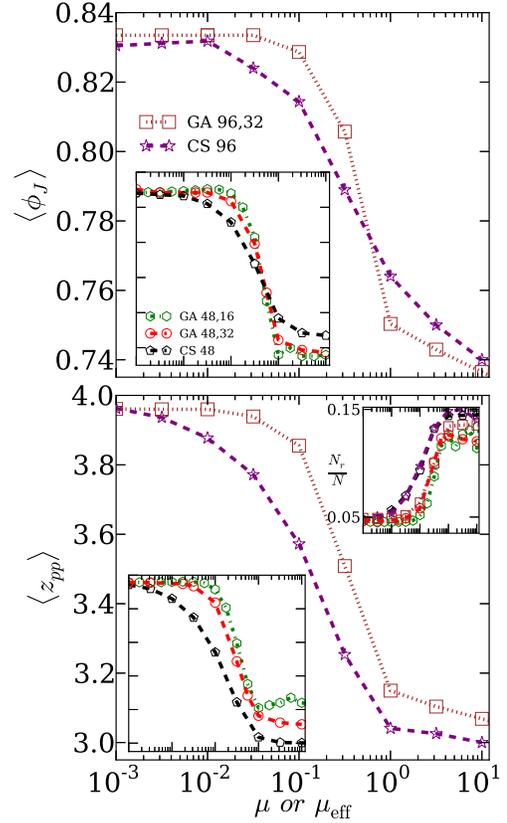}
\caption{Top: Average packing fraction $\langle \phi_J \rangle$
for MS packings from the CS and GA models versus $\mu$ or $\mu_{\rm
eff}$. The lower left inset shows $\langle \phi_J \rangle$
versus $\mu$ or $\mu_{\rm eff}$ for several system sizes $N$
and asperity numbers $N_a$. Legends show $N_a$(left) and $N$(right), and axes without tick labels
are the same as in the main panel. Bottom: Average
interparticle contact number $\langle z_{pp} \rangle$ versus $\mu$ or $\mu_{\rm eff}$. The insets show the $N$ and $N_a$ dependence of $\langle z_{pp}
\rangle$ (lower left) and rattler particle fraction $N_r/N$ (upper right).}
\label{fig:2}
\end{figure}

We construct MS packings of $N$ rough bidisperse disks ($50-50$ by
number with diameter ratio $r=1.4$) in $d=2$ using the GA model and
compare them to those from the CS approach.  The lower right panel of
Fig.~\ref{fig:1} shows rough circular disks in the GA model,
characterized by $N_a$ circular asperities with
centers on the disk rim and ratio of the asperity to particle
radius $R_a/R$.  We consider two disk interactions:
1) asperities on disks $i$ and $j$ and 2) the core of $i$ with an
asperity on $j$.  All interactions are purely repulsive linear
springs~\cite{OHern:2003kl}. Asperities $a$ and $a'$ on disks $i$ and
$j$ interact through $V^{aa'}_{ij} = \epsilon/(2 \sigma_{ij}^{2})
(\sigma^{aa'}_{ij} - r^{aa'}_{ij})^2 \Theta(1 -
r^{aa'}_{ij}/\sigma^{aa'}_{ij})$, where $r_{ij}^{aa'}$ is
the center-to-center separation between asperities,
$\sigma^{aa'}_{ij}=R^a_i+R^{a'}_j$ and $\sigma_{ij} =
\sigma_{ij}^{aa'} + R_i + R_j$.  We locate asperity $a$ on the
rim of disk $i$ at angle $\tht_{i}^{a}=\tht_i + \frac{2\pi a}{N_a}$
and coordinates ${\bf r}_{i}^{a}= {\bf r}_i + R_i(\cos\tht_{i}^{a},  \sin\tht_{i}^{a} )$, where ${\bf r}_i$ is the position of
disk $i$.  Asperity $a$ on disk $i$ and core of $j$ interact
through $V^a_{ij}= \epsilon/(2\sigma_{ij}^2) (\sigma^{a}_{ij} -
r^a_{ij})^2 \Theta(1 - r^a_{ij}/\sigma^a_{ij})$, where
$\sigma^a_{ij}=R^a_i + R_i + R_j$ (where $r_{ij}^a$ is the separation between the center of asperity $a$ on $i$ and the center of $j$).  The total GA potential energy is $V = \sum_{i>j} \sum_{a>a'} V^{aa'}_{ij} + \sum_{i>j}
\sum_a V_{ij}^a$.

\begin{figure}[bht]
\centering
\includegraphics[scale=0.4]{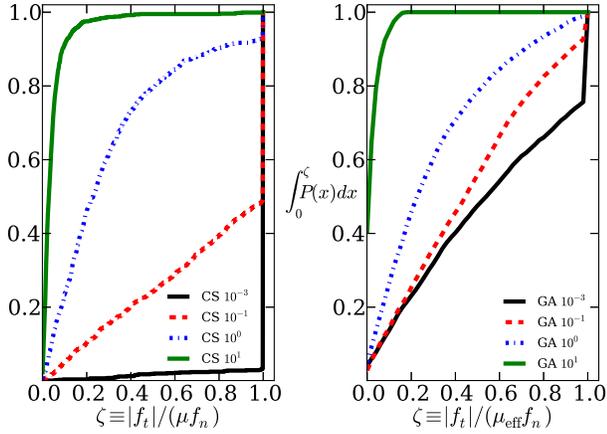}
\caption{Cumulative mobilization distributions for $N=48$ for the CS
(left) and GA (right) models for $\mu,\mu_{\rm eff}=10^{-3}$,
$10^{-1}$, $1$, and $10$, where the mobilization $\zeta = |f_t|/(\mu
f_n)$. GA mobilities $\zeta > 1$ can occur due to finite
interparticle overlaps. The bin at $\zeta = 1$ includes all $\zeta
\ge 1$ to allow a comparison with the CS model.}
\label{fig:3p5}
\end{figure}

We can define an effective GA static friction
coefficient, $\mu_{\rm eff}= 1/\sqrt{ \left((2R_a/R)/\sin
  (\pi/N_a)\right)^2 -1 }$, the maximum tangential to
normal interparticle force ratio, when an asperity on disk $i$ fits in
between two $j$'s asperities as in the lower right panel
of Fig.~\ref{fig:1}. This is the maximum 
tangential to normal force ratio in the zero interparticle
overlap limit. The ratio of the number of asperities on the large and
small particles is set close to $r$ so that the inter-species $\mu_{\rm eff}$
is approximately the same as the intra-particle one.
The CS~\cite{Silbert:2010mi,Van-Hecke:2010fv} static
friction is included between geometrically smooth circular disks $i$
and $j$ using a tangential spring with tangential to normal
spring constant ratio $k_t/k_n = 1/3$ ($k_n =
\epsilon/\sigma_{ij}$)~\cite{Shafer:1996tg}, and $|f_t|$ remains 
maximum $\mu f_n$ when $u_t$ exceeds the Coulomb
threshold. We studied system sizes from $N=6$ to $96$, 
asperity numbers $N_a = 8$, $16$, and $32$, and $\mu,\mu_{\rm eff} = 10^{-3}$ to $10$.

\begin{figure}[t!]
\centering
\includegraphics[scale=0.4]{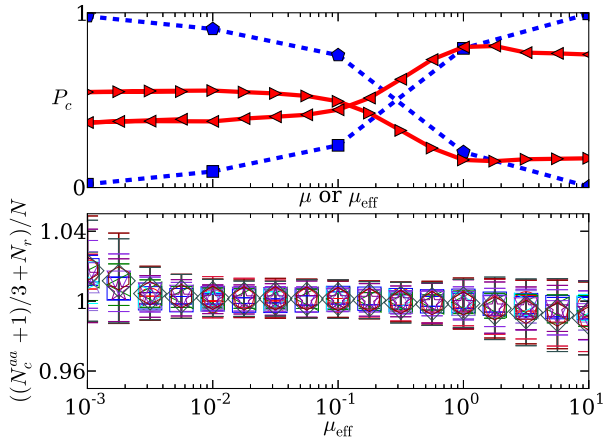}
\caption{Top: Probability $P_c$ of different contact
types versus $\mu$ or $\mu_{\rm eff}$.  
Contacts can be single
(rightward triangles) and double asperity (leftward triangles) or
low (squares) and high mobilization (pentagons) with $\zeta < \zeta_c =
0.5$ and $\zeta \ge \zeta_c$, respectively for GA (solid red) and CS models (blue dashed).  
Bottom: The average isostaticity parameter $\langle \alpha \rangle = \langle (
(N_c^{aa} + 1)/3 + N_r)/N \rangle$, where $N_{c}^{aa}$ the total
 asperity contacts, versus $\mu_{\rm eff}$ for
several $N$ and $N_a$.  ($N=6$ and $N_a = 16$, circles; $6$ and
$32$, squares; $12$ and $16$, rightward triangles; $12$ and $32$,
leftward triangles; $24$ and $16$, upward triangles; $24$ and $32$,
downward triangles; $48$ and $16$, stars; $48$ and $32$, hexagons.)
$\alpha = 1$ indicates an isostatic number of asperity contacts.}
\label{fig:3}
\end{figure}

We generate approximately $10^5$ MS GA and CS packings at
jamming onset, for each $N$ and $\mu$ or $\mu_{\rm eff}$, using the
compressive-quench-from-zero-density simulation
protocol~\cite{Gao:2006kx}. We randomly place point-particles in a
square periodic cell of unit size. We increase particle radii in small
steps corresponding to $\Delta\phi = 10^{-4}$. After each $\Delta\phi$
increment, the system is relaxed to the nearest local potential energy
minimum using dissipative forces proportional to the disks'
translational and angular velocities with large damping coefficients.
If after minimization we have zero total potential energy
per particle ({\it i.e.} $V/N < V_{\rm tol}/\epsilon = 10^{-14}$),
we keep compressing the system. Otherwise, if $V/N\ge V_{\rm tol}/\epsilon$ we decompress. $\Delta\phi$ is halved each time we switch from compression to
decompression or vice versa. 
We stop when $V_{\rm tol} < V/N < 1.01 V_{\rm tol}$, and the
average particle overlap is less than $10^{-7}$.  All GA packings are
mechanically stable with $3N' - 2$ eigenvalues $m_i>0$ for the
dynamical matrix $M_{kl} = \frac{d^2 V}{d {\bf R}_k d {\bf R}_l}$,
where ${\bf R} = \{ {\bf r}_1,\ldots,{\bf r}_{N'},
(R_1+R^a_1)\theta_1,\ldots,(R_{N'}+R^a_{N'})\theta_{N'}\}$,
$N'=N-N_r$, and $N_r$ the rattler particles. 
(CS and GA rattler particles have less than three
interparticle contacts) Fig.~\ref{fig:2} shows results for the average packing fraction
$\langle \phi_J \rangle$ and contact number $\langle
z_{pp} \rangle = \langle 2N_{pp}/(N') \rangle$ at jamming onset,
where $N_{pp}$ the particle-particle contacts irrespective of the number of asperity contacts. 
As previously~\cite{Silbert:2010mi}, $\langle \phi_J
\rangle$ varies from $\approx 0.84$ to $0.75$ and $\langle z_{pp}
\rangle$ ranges from $\approx 4$ to $3$ as $\mu$
increases for both CS and GA models.  The crossover from frictionless to frictional behavior
occurs near $\mu^* \approx 0.1$. $\langle \phi_J \rangle$ is
$1\%$ larger at large $\mu_{\rm eff}$ for the GA model, 
expected for finite $N_a$.  The upper right panel of Fig.~\ref{fig:2}
shows $N_r/N$ versus $\mu$ or $\mu_{\rm eff}$.  Both increase with $\mu$ or $\mu_{\rm eff}$
and then plateau.  Due to slow relaxation processes we
detect fewer rattlers for the GA model, causing $\langle z_{pp}
\rangle$ to be $5\%$ larger at large $\mu_{\rm eff}$.

The cumulative mobilization distributions ($A(\zeta) = \int_0^{\zeta}
P(x) dx$, where $\zeta = |f_t|/(\mu f_n)$) are qualitatively similar
for the CS and GA models in Fig.~\ref{fig:3p5}. At low $\mu$ or
$\mu_{\rm eff}$, $A(\zeta)$ for both models has a strong peak at
$\zeta = 1$~\cite{Silbert:2002ij,Shundyak:2007fk}.  As $\mu$ or
$\mu_{\rm eff}$ increases, it disappears and the average
mobilization decreases. Quantitative differences in
the mobilization distributions are due to the different tangential force
laws shown in the lower left panel of Fig.~\ref{fig:1}. At fixed
overlap, $f_t/f_n$ varies linearly with $u_t$ until the sliding limit
at $\pm u_a$, while $f_t/f_n$ is periodic for the GA model.  

\begin{figure}[t!!]
\centering
\includegraphics[scale=0.4]{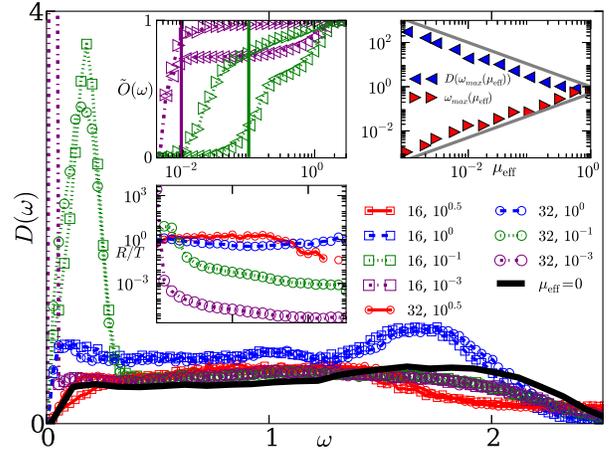}
\caption{$D(\omega)$ for $N=48$, $N_a = 16$ and $32$, and $\mu_{\rm
eff} = 10^{-3}$, $10^{-1}$, $1$, and $10^{0.5}$ for the GA model.
The area under $D(\omega)$ for $\mu_{\rm eff} >0$ is the number of
nonzero modes $3N'-2$, while for $\mu_{\rm eff}=0$ is
$2N'-2$. 
The upper-left inset shows  $\tilde O(\omega)$ after simple
shear (rightward triangles) and compression (leftward triangles) perturbations at high and low
friction (different colors in main legend). Vertical solid lines indicate corresponding locations of the $D(\omega)$ peak. 
Lower-left inset shows $R/T$ for the modes on a linear frequency
scale. The upper right inset tracks $\omega_{\rm max}$
and $D(\omega_{\rm max})$ for the low-frequency peak in
$D(\omega)$ versus $\mu_{\rm eff}$ for $N_a=32$.  Solid lines
have slopes $-1$ and $1$.}
\label{fig:4}
\end{figure}

In the lower panel of Fig.~\ref{fig:3}, we show the
asperity contacts (single, double, and triple) for each interparticle
contact.  We find that MS packings are {\it isostatic}~\cite{Donev2005} with $N_c^{aa}
= 3N' - 1$ contacts over the entire range of $\mu_{\rm eff}$.
Deviations from isostaticity are less than $2\%$ for all $N$ and $N_a$
studied.  In contrast, static packings of frictional particles are
hyperstatic ($z_{pp} > 3$) when considering interparticle contacts for
both GA and CS~\cite{Van-Hecke:2010fv} ({\it cf.} lower panel of
Fig.~\ref{fig:2}).

Asperity contacts may explain the structural and mechanical crossover near $\mu^*$.  
In the top panel of Fig.~\ref{fig:3},
we plot the probability of single and double asperity contacts versus
$\mu_{\rm eff}$.  They are roughly
equiprobable at low friction, while only double asperity contacts
occur at high friction. To maintain isostaticity, at low friction
there are typically two double and two single asperity contacts per
particle, while at high friction three double contacts form for a
total of approximately six per particle in both cases.  The $\mu_{\rm
  eff}$ where single become less probable than double asperity
contacts ($\sim0.1$) coincides with $\mu^*$ above which the packing
fraction, contact number, and mobilization distributions begin to deviate
significantly from frictionless behavior. 
Such competition
also occurs for the CS model. In the upper panel of
Fig.~\ref{fig:3}, we show the probability of low ($\zeta < \zeta_c =
0.5$) and high ($\zeta \ge \zeta_c$) mobilization contacts versus $\mu$.
(The results do not depend strongly on $\zeta_c$.)  At
low friction, most contacts possess high mobilization, while they have low mobilization at high friction.  At high friction,
double asperity contacts resemble low mobilization contacts.  At low friction, both single and double asperity contacts can
possess high mobilization.  The crossover in the
probabilities of low and high mobilization contacts occurs also near $\mu^*$.

We can directly calculate the GA 
 $D(\omega)$ from the total potential
energy (in the harmonic approximation). The eigenmode with frequency $\omega_j$ is ${\hat{\bf m}}_j = \{
m^{x,1}_j, m^{y,1}_j, m^{\theta,1}_j,\ldots,m^{x,N'}_j, m^{y,N'}_j,
m^{\theta,N'}_j \}$ with $\sum_{\lambda,i}(m_j^{\lambda,i})^2=1$.  The rotational $R_j$ and translational $T_j$ content of each
mode $j$ are $T_j = \sum_{i=1,N'} \sum_{\lambda=x,y} (m^{\lambda
i}_j)^2$, and $R_j = 1 - T_j$; the participation ratio $P_j =
(\sum_{\lambda,i} (m^{\lambda,i}_j)^2)^2/(N \sum_{\lambda,i}
(m_j^{\lambda,i})^4)$ for $\lambda = x,y$ and $\theta$ separately, and
the optical order parameter $Q^{\rm opt}_j = \sum_{i,k} m_j^{\theta,
i} m_j^{\theta, k}/(N \sum_i (m_j^{\theta, i})^2)$ that characterizes
whether the rotational content of $j$ is co- or
counter-rotating~\cite{Henkes:2010qa}.

$D(\omega)$ for MS packings using the GA model is shown in
Fig.~\ref{fig:4}: (i) A
strong peak at low frequency whose height $D(\omega_{\rm max})$
increases and location $\omega_{\rm max}$ shifts to lower frequency
with decreasing $\mu_{\rm eff}$. We find that $\omega_{\rm max} \sim
\mu_{\rm eff}$ and $D(\omega_{\rm max}) \sim \mu_{\rm eff}^{-1}$ as
$\mu_{\rm eff} \rightarrow 0$ ({\it cf.} upper-right inset of
Fig.~\ref{fig:4}). These modes are mostly rotational ($R \sim 1$),
globally incoherent ($Q^{\rm opt} \sim 0$), and quasi-localized
($P\lesssim 0.1$) as $\mu_{\rm eff}\rightarrow 0$.  Similar peaks in
$D(\omega)$ that contain low-frequency rotational modes have been
 found in ellipse packings~\cite{Schreck:2010zr,Schreck:2012dq} at low aspect ratio.  For small $\mu_{\rm eff}$, as $\omega$ increases,
$D(\omega)$ approaches the frictionless case with
 translational and quasi-localized modes at high
frequencies.  (ii) A peak in $D(\omega)$ at low frequency with
$R\sim 1$ disappears for $\mu_{\rm eff} \gtrsim \mu^*$.  (iii) For
$\mu_{\rm eff} \gtrsim \mu^*$, modes have mixed rotational and
translational content with $R\sim T$ at all frequencies.  At low
frequencies, modes are
``gear-like''~\cite{Baram:2004bh,Herrmann:1990ve,Oron:2000qf} ($Q_{\rm
  opt} \sim -0.5$) and collective ($P \sim 0.3$).  At high
frequencies, modes are increasingly localized with co-rotating
angular components ($Q_{\rm opt} \sim 0.5$).

Low-frequency rotational
modes couple strongly to the mechanical response of GA packings, shown by quasistatic a) isotropic compression in packing
fraction increments to total $\Delta \phi_{\rm tot}
=10^{-8}$ or b) simple shear in strain increments (coupled with
Lees-Edwards boundary conditions) to $\gamma_{\rm tot}
= 10^{-8}$ from a reference configuration at $\Delta \phi_0
= 10^{-6}$.
 We calculated the overlap
$O(\omega)=\delta{\bf D} \cdot\hat{\bf m}_j(\omega)/|\delta{\bf D}|^2$ of the deformation
vector $\delta{\bf D} \equiv {\bf D} - {\bf D}_0$, where ${\bf D}_0$
(${\bf D}$) is the $3N'$-dimensional coordinate vector of the
reference configuration.  In the upper inset of
Fig.~\ref{fig:4}, the low-frequency rotational modes
contribute to at least half of the cumulative and averaged absolute overlap $\tilde
O(\omega)=\int_{0}^{\omega}|O(\omega')|d\omega'/\int_{0}^{\infty}|O(\omega')|d\omega'$
for both compression and shear.

\begin{acknowledgements}
Support from NSF Grant No. CBET-0968013 (M.S.) and DTRA Grant
No. 1-10-1-0021 (S.P. and C.O.) is acknowledged. This work also
benefited from the facilities and staff of the Yale University Faculty
of Arts and Sciences High Performance Computing Center and NSF Grant
No. CNS-0821132 that partially funded acquisition of the computational
facilities. We thank T. Bertrand, K. Kumar, D. Kwok, M. Wang, and
C. Schreck for their invaluable insights and comments throughout the 
course of this work.  
\end{acknowledgements}


\end{document}